\documentclass[twocolumn,pra,showpacs,preprintnumbers,floatfix,amsmath,amssymb]{revtex4}

\usepackage{graphicx}
\usepackage{dcolumn}
\usepackage{bm}

\begin{document}
\newcommand{\ket}[1]{|#1\rangle}
\newcommand{\bra}[1]{\langle#1|}
\preprint{APS/123-QED}

\title{Experimental observation of transient velocity-selective coherent population trapping in one dimension }

\author{Frank Vewinger}
\email{vewinger@physik.uni-kl.de}
\author{Frank Zimmer}%
 \email{zimmer@physik.uni-kl.de}
\affiliation{%
Fachbereich Physik\\
Technische Universit{\"a}t Kaiserslautern\\
D-67663 Kaiserslautern
}%

\date{\today}
%
%
%
\begin{abstract}
We report the observation of transient velocity-selective coherent
population trapping (VSCPT) in a beam of metastable neon atoms.
The atomic momentum distribution resulting from the interaction
with counterpropagating $\sigma_+$ and $\sigma_-$ radiation which
couples a $J_g=2\leftrightarrow J_e=1$ transition is measured via
the transversal beam profile. This transition exhibits a stable
VSCPT dark state formed by the two $|J=2,m=\pm1>$ states, and a
metastable dark state containing the $|J=2,m=\pm 2>$ and
$|J=2,m=0>$ states. The dynamics of the formation and decay of
stable and metastable dark states is studied experimentally and
numerically and the finite lifetime of the metastable dark state
is experimentally observed. We compare the measured distribution
with a numerical solution of the master equation.
\end{abstract}

\maketitle
%
%

Velocity-selective coherent population trapping (VSCPT) occurs by
optical pumping of atoms into states with well defined momenta,
which are decoupled from the light field. In a
$J_g=1\leftrightarrow J_e=1$ transition  a dark state exists,
which is a trapping state for zero momentum, leading to efficient
cooling below the one-photon recoil limit
\cite{Aspect88,Aspect89}, which may be assisted by
polarization-gradient precooling \cite{Shahriar93}. For ground
state angular momenta $J_g>1$ several dark states may exist. These
are not necessarily eigenstates of the kinetic energy Hamiltonian
and thus they are transient \cite{Papoff92}, not leading to
population trapping. The stability of those states may be
recovered by introducing $m$-dependent shifts of the Zeeman
sublevels by a dc stark field \cite{Olshanii91} or additional
laser fields \cite{Menotti97}. The extension of VSCPT on three
dimensions has also been discussed \cite{Olshanii92}. The
existence of transient dark states has for instance been used to
demonstrate a multiple beam atomic interferometer in cesium
\cite{Weitz96}. Furthermore, for specific polarizations of the
laser fields the existence of high-velocity dark states has been
shown \cite{Widmer96}. Recently studies of velocity-selective
coherent population trapping and its relation to
electromagnetically induced transperency (EIT) and atomic
entanglement where reported \cite{Kiffner05}.

For a $J_g=2\leftrightarrow J_e=1$ transition coupled by
counterpropagating $\sigma_+$ and $\sigma_-$ beams the existence
of transient VSCPT has been shown theoretically \cite{Papoff92},
but to the best of our knowledge experimental momentum
distributions have not been reported. In particular the
metastability of some of the dark states has not been directly
shown in the experiment. We present measurements that clearly show
the signature of a stable and a transient dark state in the
momentum distribution of a beam of neon atoms in the metastable
state $^3P_2$.

\begin{figure}
\includegraphics[width=\columnwidth]{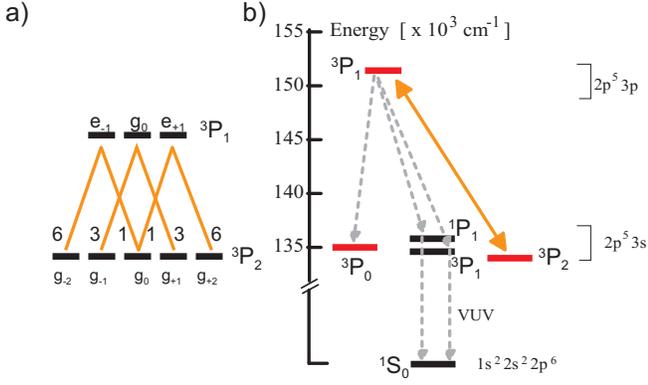}
\caption{\label{coupling}a) Coupling scheme for the
$J=2\leftrightarrow J=1$ transition driven by $\sigma_+$ and
$\sigma_-$ polarized light. The numbers are the squares of the
ratios of the Clebsch-Gordan coefficients. b) Level scheme of neon
including the relevant levels for the experiment. }
\end{figure}

The coupling scheme for the $J_g=2\leftrightarrow J_e=1$ transition
driven by $\sigma_+$ and $\sigma_-$ irradiation with equal frequency is shown in Figure
\ref{coupling}. The system is characterized by the Hamiltonian,
\begin{equation}
\begin{split}
H&=\frac{{\bf p}^2}{2M}+\hbar \omega_0 P_e+V_{AL},
\end{split}
\end{equation}
where $P_e$ is the projector onto the excited states, ${\bf p}$
the momentum operator and $M$ is the atomic mass. The
interaction Hamiltonian $V_{AL}=V_{\Lambda}+V_{IW}$ consists of
two parts. One describes the $\Lambda$-system
$\{g_{-1},e_0,g_{+1}\}$, the other the inverted-W configuration
$\{g_{-2},e_{-1},g_{0},e_{+1},g_{+2}\}$. In the rotating wave
approximation the two terms read
\begin{eqnarray}
V_{\Lambda}&=&\sum_q\frac{\hbar}{2}{\textstyle\sqrt{\frac{3}{10}}}\{\Omega_{+}\ket{e_0,q}\bra{g_{-1},q-\hbar k}\\
&+&\Omega_{-}\ket{e_0,q}\bra{g_{+1},q+\hbar k}\}\exp(-i
\omega_L t) + h.c.,\nonumber\\
V_{IW}&=&\sum_q\frac{\hbar}{2}\left[{\textstyle\sqrt{\frac{6}{10}}}\Omega_{+}\ket{e_{-1},q-\hbar
k}\bra{g_{-2},q-2\hbar k}\right.\\
&+&\!\!\!{\textstyle\sqrt{\frac{1}{10}}}\big[\Omega_{-}\ket{e_{-1},q-\hbar
k}
    +\Omega_{+}\ket{e_{+1},q+\hbar k}\big]\bra{g_{0},q}\nonumber\\
&+&\!\!\!\left.{\textstyle\sqrt{\frac{6}{10}}}\Omega_{-}\ket{e_{+1},q+\hbar
k}\bra{g_{+2},q+2\hbar k}\right]e^{-i \omega_L t}\! +
h.c.,\nonumber
\end{eqnarray}
where $\Omega_{+}$ ($\Omega_{-}$) is the Rabi frequency of the
coupling laser for the transition  $\ket{g_0}\leftrightarrow
\ket{e_{+1}}$ ($\ket{g_0}\leftrightarrow\ket{e_{-1}}$).

The interaction Hamiltonian $V_{AL}$ separates the atomic states
into two sets of momentum families, which are coupled to each
other only by spontaneous emission. These are
\begin{eqnarray}
 \mathfrak{F}^{\Lambda}(q)&=&\{\ket{e_0,q},\ket{g_{-1},q-\hbar k},\ket{q_{+1},q+\hbar k}\}\\
 \mathfrak{F}^{IW}(q)&=&\{\ket{g_{-2},q-2\hbar k},\ket{e_{-1},q-\hbar
 k},\ket{g_0,q},\nonumber\\
 &&\ket{e_{+1},q+\hbar k},\ket{g_{+2},q+2\hbar k}\}.
\end{eqnarray}
The $\mathfrak{F}^{\Lambda}(q)$ family involves two states which
have for $q=0$ the same energy. Thus two-photon resonance within
the $\Lambda$-system can be maintained using a single laser
frequency and the dark state $\ket{\Psi_{NC}^{\Lambda}(q=0)}$
formed by the members of this family will be stable. The
$\mathfrak{F}^{IW}(q)$ family involves states with different
momenta. Thus two-photon resonance between the states
$\ket{g_{\pm2}}$ and $\ket{g_0}$ can not be maintained with a
single laser frequency. Furthermore the dark state
$\ket{\Psi_{NC}^{IW}}$ formed within this family is no eigenstate
of the kinetic energy Hamiltonian ${{\bf p}^2}/{2M}$. The lifetime
$\tau_{IW}$ of this state has been calculated perturbatively,
assuming that all Clebsch-Gordan coefficients are equal
\cite{Papoff92}. The two dark states read
\begin{eqnarray}
\ket{\Psi_{NC}^{\Lambda}}&=&{\frac{1}{\sqrt{\Omega^2_{+}+\Omega^2_{-}}}}\nonumber\\
&\times&
\big[\Omega_{-}\ket{g_{-1},q-\hbar k}-\Omega_{+}\ket{g_{+1},q+\hbar k}\big],\label{dark_state_lambda}\\
\ket{\Psi_{NC}^{IW}}&=&\frac{1}{\sqrt{\Omega^4_{+}+6\Omega^2_{+}\Omega^2_{-}+\Omega^4_{-}}}\Big[\Omega^2_{-}\ket{g_{-2},q-2\hbar
k}\nonumber\\
&&-\sqrt{6}\Omega_{+}\Omega_{-}\ket{g_0,q}+\Omega^2_{+}\ket{g_{+2},q+2\hbar
k}\Big].\label{dark_state_IW}
\end{eqnarray}
For interaction times $\tau<\tau_{IW}$ both dark states appear as
trapping states for $q=0$, giving rise to a momentum distribution
with peaks at $-2\hbar k,-\hbar k,0,\hbar k$ and $2\hbar k$. For
interaction times $\tau>\tau_{IW}$ the contribution of
$\ket{\Psi_{NC}^{IW}}$ vanishes and only two peaks at $\pm \hbar
k$ remain in the momentum distribution.


\begin{figure}
\includegraphics[width=6cm]{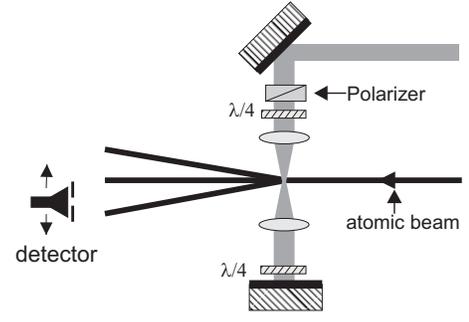}
\caption{\label{setup}Schematic setup of the experiment. The
cylindrical lenses have a focal length of 250 mm and they are in a
confocal arrangement. Further details can be found in the text.}
\end{figure}
In the experiment, a beam of neon atoms emerges from a liquid
nitrogen cooled discharge nozzle source. A fraction of the order
of $10^{-4}$ of the atoms is in the metastable states ${}^3\!P_0$
or ${}^3\!P\/_2$ of the $2p^53s$ electronic configuration
\cite{weber98}. The flow velocity of the atoms is about
$470$~ms${}^{-1}$ with a width of the velocity distribution of
about $100$~ms${}^{-1}$ (FWHM). The beam is collimated by a 50
$\mu$m and a 10~$\mu$m slit positioned 144 cm apart, corresponding
to a width of the transversal velocity distribution of about
$0.15\hbar k$. The population of the metastable ${}^3\!P_0$ state
is depleted by optical pumping before the atoms interact with the
circular polarized laser beams in a region 20 cm downstream of the
second slit. The spatial distribution of the atoms in the states
${}^3\!P_0$ and ${}^3\!P_2$ is measured 120~cm downstream of the
interaction zone using a movable channeltron detector behind a
entrance slit of 25~$\mu$m width, {leading to a resolution of the
transverse momentum of  $\Delta p\approx 0.2\hbar k$}. In the
interaction region the magnetic field is actively compensated to
less than 1~$\mu$T, to assure the degeneracy of all Zeeman states
to within better than 130~kHz. The laser beam passes through a
polarizer, two quarter wave plates and optionally two cylindrical
lenses before being retroreflected (Fig.\ \ref{setup}),
establishing a counterpropagating
$\sigma_+-\sigma_-$-configuration. {In order to realize different
interaction times three different setups for the width of the
laser beams were used.} The first setup uses cylindrical lenses in
a confocal arrangement with the atomic beam crossing the laser
beam near the focus. The transit time of the atoms through the
laser beam is estimated to be a few 100~ns, corresponding to
$\Gamma t\approx 10$, where $\Gamma$ is the width of the
transition between the ${}^3\!P_1$ and ${}^3\!P\/_2$ state. The
laser beam profile at the position of the atomic beam is not
measured directly but inferred from the dimensions of the optical
setup. The laser beams are parallel to within $10^{-5}$~rad. The
peak Rabi frequency is in the order of 500~MHz and the lasers are
tuned from the $\ket{g_{0}}\leftrightarrow\ket{e_0}$ resonance by
100~MHz to reduce the influence of stray light from the windows.
To increase the interaction between the atoms and the light field
the cylindrical lenses can be removed, leading to an interaction
time of $\Gamma t=200$. Using a telescope in front of the
polarizer the beam diameter can be increased to 8~mm, leading to
an interaction time of $\Gamma t=800$.



\begin{figure}
\includegraphics[width=\columnwidth]{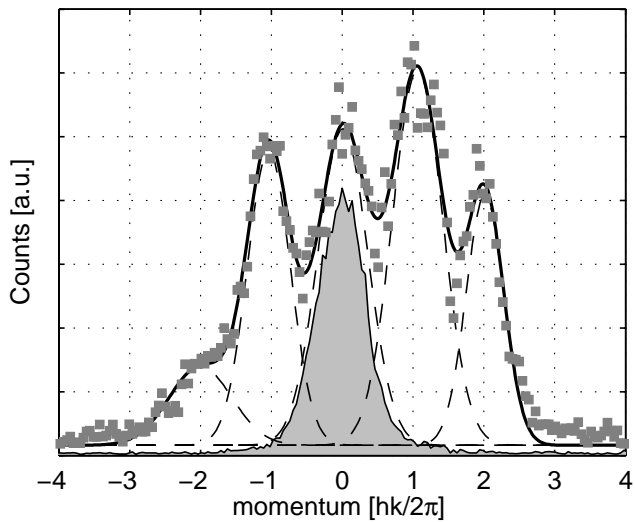}
\caption{Transverse atomic momentum profile after a short
interaction ($\Gamma t<10$). The grey area shows the initial
momentum distribution (scaled down), the dashed lines are gaussian
fits to the individual peaks. The full line is the sum of the
gaussian fits.}\label{results}
\end{figure}

{Figure \ref{results} shows the initial momentum distribution
(grey area) and the result for a short interaction time of $\Gamma
t\approx10$ (squares)}. Five peaks at $\pm2 \hbar k, \pm \hbar k$
and zero momentum are clearly resolved.
The asymmetry of the momentum distribution is due to different
intensities of the $\sigma_+$ and $\sigma_-$ beams: The
retroreflected beam passes twice through a window and the
(uncoated) cylindrical lens before crossing the atomic beam,
resulting in a lower intensity in the retroreflected beam. {Thus
the Rabi frequencies of the two laser beams are not equal,
$\Omega_{+}\neq\Omega_{-}$, therefore we find an asymmetric
population distribution within the dark states
(\ref{dark_state_lambda}) and (\ref{dark_state_IW}).} This is also
confirmed by numerical simulations of the process. The peak at
zero momentum also contains contributions from population in the
$^3P_0$ state, which is populated by spontaneous emission from the
upper state $^3P_1$ during the interaction.

\begin{figure}
\includegraphics[width=\columnwidth]{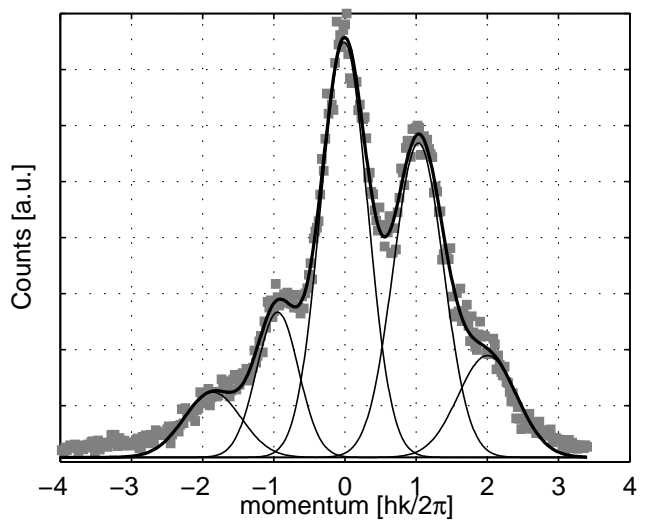}
\caption{\label{asymm}Transversal momentum distribution after the
interaction with a tilted retroreflected laser. The dashed lines
are gaussian fits to the individual peaks, the solid line is their
sum. The depopulation of the peaks with negative momentum is
clearly visible.}
\end{figure}

{In order to do a supplementary test that we observe
velocity-selective coherent population trapping the retroreflected
beam was slightly tilted . A sufficient overlap of both beams was
sustained but the retroreflected beam interacts with the atoms
after the dark states have been populated.} Due to optical pumping
the population of the states $\ket{g_{-2}}$, $\ket{g_{-1}}$ and
$\ket{g_{0}}$ is depleted, and the height of the peaks at negative
momenta decrease, since the internal and external states in the
dark states are strongly correlated. The measurement with tilted
lasers is shown in figure \ref{asymm}, which shows good agreement
of experimental and calculated profiles.

\begin{figure}
  \includegraphics[width=\columnwidth]{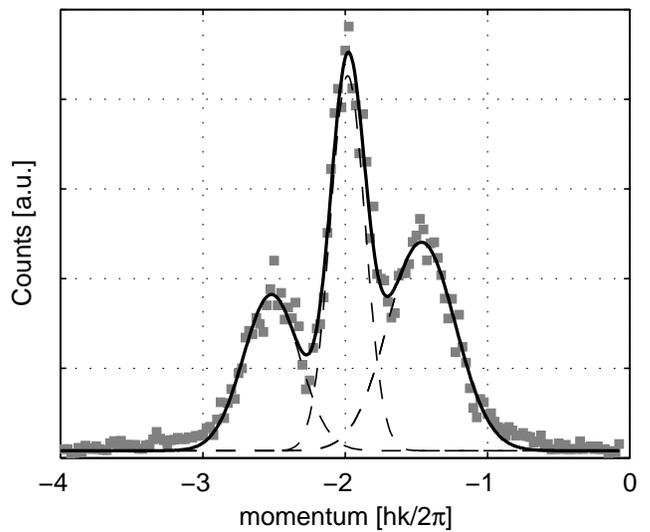}\\
  \caption{Transverse atomic momentum profile after an
interaction of 8~$\mu$s ($\Gamma t\approx 800$). The lines are
gaussian fits to the peaks. The peaks at $p=\pm \hbar k$ reflect the stable dark-state
$|\Psi_{\rm NC}^{\Lambda}(q=0)\rangle$.}\label{interaction_long}
\end{figure}

When increasing the laser beam diameter with a telescope to 8~mm
the interaction time is of the order $\Gamma t\approx 800$. We
then observe the transversal beam profile shown in figure
\ref{interaction_long}. The peaks at $\pm 2\hbar k$ are no longer
seen and only the stable dark state $\ket{\Psi_{NC}^{\Lambda}}$
survives, reflected by the peaks at the {momenta} $\pm \hbar k$.
The peak at zero momentum appears because the state $^3P_0$ is
populated by spontaneous emission during the process of dark state
preparation.


\begin{figure}
  \includegraphics[width=\columnwidth]{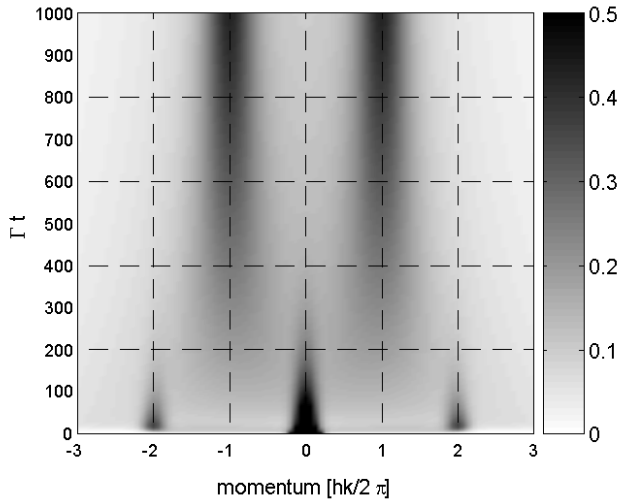}\\
  \caption{Dynamics of the momentum distribution, derived from a numerical solution
  of the Bloch equations (\ref{master-eqn}). The population is encoded by the greyscale
  given on the right. }\label{pcolor_plot}
\end{figure}
We compare the measured data to a solution of the generalized
optical Bloch equations in the family momentum
basis \cite{Aspect89}, neglecting spontaneous decay to other levels
outside the system. The Bloch equation reads
\begin{eqnarray}
\dot\rho&=&-\frac{i}{\hbar}\left[ H,\rho \right]\nonumber\\
&-&\frac{\Gamma}{2}\left[\left(\Delta^{+}\cdot\Delta^{-}\right)
   \rho+\rho\left(\Delta^{-}\cdot\Delta^{+}\right)\right]
   \nonumber\\
&+&\frac{3\Gamma}{8\pi}
   \int{\rm d}^2\Omega\sum
   \limits_{{\bf\epsilon}\bot{\bf n}}
   \left(\Delta^{+}\cdot{\bf\epsilon}\right)^{\dagger}
   \exp(-ik{\bf n}\cdot{\bf R})\rho\label{master-eqn}\\
&~&   \times\exp(ik{\bf n}\cdot{\bf R})
   \left(\Delta^{+}\cdot{\bf\epsilon}\right),\nonumber
\end{eqnarray}
{where $\Delta_{\pm}$ are the lowering and raising parts of the
reduced dipol operator and ${\bf\epsilon}$ is the polarization
vector \cite{Castin89}. ${\bf R}$ is the position operator of the
center of mass which acts only on the external variables.} The
second term describes the decrease of the excited-state
populations and coherences resulting from spontaneous emission;
the term under the integral over all solid angles (third term)
describes the feeding of the ground-state population and
coherences by spontaneous emission of a fluorescence photon into
the solid angle around the direction ${\bf n}$, with energy $\hbar
ck$ and polarization ${\bf\epsilon\bot n}$. The explicit form of
the differential equations for the density matrix elements are
given in \cite{wallis95}. The equations were integrated stepwise
with a resolution of $1/50 \Gamma^{-1}$ in time and $\hbar k/20$
in momentum space on an interval of $[-8 \hbar k,8 \hbar k]$. The
calculations were done for a Rabi frequency of $0.3\Gamma$ with no
time dependence. The initial momentum distribution is given by a
gaussian profile of width $\Delta q=0.15\hbar k$ centered at
$p=0$, which is taken from the experiment. Initially all Zeeman
states in the $^3P_2$ manifold are equally populated. The results
are shown in Fig.\ \ref{pcolor_plot}.

For a short interaction time of $\Gamma t<15$, the atoms absorb a
photon from one of the laser beams, followed by stimulated
emission into the other beam. This leads to a change in momentum
by $\Delta p=\pm2\hbar k$, and peaks in the momentum distribution
at $p=\pm2\hbar k$ appear. This process of stimulated raman
transitions causes an efficient population transfer to the dark
state $\ket{\Psi_{\rm NC}^{IW}(q=0)}$, characterized by a momentum
distribution located at $p=\pm 2 \hbar k$ and $p=0$. Absorption of
a photon followed by spontaneous emission {results in} a random
walk in momentum space due to the emission of photons in {an
arbitrary} direction. This diffusive process successively
populates the stable dark state $\ket{\Psi_{\rm
NC}^{\Lambda}(q=0)}$ which is characterized by the two peaks at
momentum $p=\pm\hbar k$ in the momentum representation. Due to the
statistical nature of this process the population of the dark
state $\ket{\Psi_{\rm NC}^{\Lambda}(q=0)}$ increases for longer
interaction times, $\Gamma t>200$.


The calculations show good agreement with the measured data. The
calculations show a structure with five peaks at $-2\hbar
k,...,2\hbar k$ for $\Gamma t\approx 100-200$, while the
measurements yield this structure for $\Gamma t\approx 10-20$. The
values for the interaction time are not directly comparable, as
the calculations where done for a constant Rabi frequency, while
in the experiment the light fields have a gaussian shape.
Furthermore the widths of the measured peaks is smaller than
expected from the simulations, which is due to neglecting
spontaneous emission out of the system $\{^3P_1,{}^3P_2\}$.

The measurements for a long interaction time of $\Gamma t=800$
(Fig. \ref{interaction_long}) show good agreement with the
numerical results. The population of the stable dark state
(\ref{dark_state_lambda}) is rising, while the population of other
states is decaying into this dark state via the diffusion in
momentum space due to the spontaneous emission of photons.

In this work we {have} presented measurements which directly
demonstrate the existence of transient dark states with a well
defined momentum distribution in a $m$-state manifold of a
$J_g=2$-level coupled to a level with $J_e=1$ by
counterpropagating $\sigma_+$ and $\sigma_-$ radiation. The
measured data show good agreement with quantum density matrix
calculations for the velocity distribution for short as well as
longer interaction times. For a quantitative analysis more
detailed experiments as well as calculations including spontaneous
emission into other states than the $^3P_2$-state are needed.

We thank K. Bergmann and M. Fleischhauer for their support,
discussions and helpful comments on the manuscript. We thank M.
Heinz for his contribution to the experiments. This work was
supported by the Deutsche Forschungsgemeinschaft (Project Be
623/32) and under the Graduiertenkolleg 792 'Nichtlineare Optik
und Ultrakurzzeitphysik'.

\bibliography{vscpt_ver1}

\end{document}